\documentclass[12pt]{article}
\usepackage{graphicx}
\usepackage{subfigure}
\usepackage{color}

\newcommand\pubnumber{SNSN-323-63}
\newcommand\pubdate{\today}

\def\napoli{Institute of Physics, National Chiao-Tung University, Hsinchu}

\def\Title#1{\begin{center} {\Large #1 } \end{center}}
\def\Author#1{\begin{center}{ \sc #1} \end{center}}
\def\Address#1{\begin{center}{ \it #1} \end{center}}

\newcommand\pubblock{\rightline{\begin{tabular}{l} \pubnumber\\
         \pubdate  \end{tabular}}}
\newenvironment{Abstract}{\begin{quotation}  }{\end{quotation}}
\newenvironment{Presented}{\begin{quotation} \begin{center} 
             PRESENTED AT\end{center}\bigskip 
      \begin{center}\begin{large}}{\end{large}\end{center} \end{quotation}}
\def\Acknowledgements{\bigskip  \bigskip \begin{center} \begin{large}
             \bf ACKNOWLEDGEMENTS \end{large}\end{center}}

\def\beq{\begin{equation}}
\def\eeq#1{\label{#1}\end{equation}}
\def\eeqn{\end{equation}}

\def\beqa{\begin{eqnarray}}
\def\eeqa#1{\label{#1}\end{eqnarray}}
\def\eeqan{\end{eqnarray}}

\let\bar=\overbar

\def\Dslash{\not{\hbox{\kern-4pt $D$}}}
\def\dslash{\not{\hbox{\kern-2pt $\del$}}}

\def\msb{{\bar{\ssstyle M \kern -1pt S}}}

\begin{document}
\begin{titlepage}
\pubblock

\vfill
\Title{New Results from the Daya Bay Reactor Neutrino Experiment}
\vfill
\Author{ Bei-Zhen Hu, for the Daya Bay Collaboration}
\Address{\napoli}
\vfill
\begin{Abstract}
This presentation describes a precision result of the neutrino mixing parameter, $\sin^2 2\theta_{13}$, and the first direct measurement of the antineutrino mass-squared difference $\sin^2(\Delta_{ee}) \equiv  \cos^2 \theta_{12} \sin^2 \Delta_{31} + \sin^2 \theta_{12} \sin^2 \Delta_{32}$ from the Daya Bay Reactor Neutrino Experiment. The above results are based on the six detector data-taking from 24 December 2011 to 28 July 2012. By using the observed antineutrino rate and the energy spectrum analysis, the results are $\sin^2 2\theta_{13}=0.090^{+0.008}_{-0.009}$ and $| \Delta m^2_{ee}| = 2.59^{+0.19}_{-0.20} \cdot 10^{-3} $eV$^2$ with a $\chi^2$/NDF of 162.7/153. The value of $| \Delta m^2_{ee}| $ is consistent with $| \Delta m^2_{\mu\mu}|$ measured in muon neutrino beam experiments.

\end{Abstract}
\vfill
\begin{Presented}
Symposium on Cosmology and Particle Astrophysics\\
Honolulu, Hawai'i,  November 12--15, 2013
\end{Presented}
\vfill
\end{titlepage}
\def\thefootnote{\fnsymbol{footnote}}
\setcounter{footnote}{0}

\section{Introduction}
The neutrino flavor eigenstates are linear combination of the mass eigenstates with the unitary matrix, Pontecorvo-Maki-Nakagawa-Sakata (PMNS) matrix. In the PMNS matrix, the three mixing angles ($\theta_{12}$, $\theta_{23}$, $\theta_{13}$ ) can be measured experimentally. The Daya Bay collaboration has published the first non-zero $\theta_{13}$ results with a significance of 5.2 standard deviations in 2012 \cite{dybPRL, dybCPC}.

The survival probability for electron antineutrino in vacuum with an energy $E$ at a distance $L$ is given by 
 \begin{equation}
	P_{ee} = 1- \cos^4 \theta_{13} \sin^2 2\theta_{12} \sin^2 \Delta_{21} - \sin^2 2\theta_{13}( \cos^2 \theta_{12} \sin^2 \Delta_{31} + \sin^2 \theta_{12} \sin^2 \Delta_{32} ),
\end{equation}
where $ \Delta_{ij} \equiv 1.267 \Delta m^2_{ij}(\rm eV^2)\frac{L(\rm m)}{E( \rm MeV)}$, and $\Delta m^2_{ji}$ is the squared-mass difference between the mass eigenstates. 

By measuring $P_{ee}$ from the reactor antineutrinos, the mixing angle $\theta_{13}$ and the oscillation function, defined as $\sin^2 \Delta_{ee} \equiv \cos^2 \theta_{12} \sin^2 \Delta_{31}+\sin^2 \theta_{12} \sin^2 \Delta_{32} $, can be determined.

In this talk, we shall present a precision measurement of $\sin^2 2\theta_{13}$ and the first measurement of $| \Delta m^2_{ee}| $ with 217 days of data at Daya Bay.

\section{Experimental Setup}
The Daya Bay experimental site is located at the southern part of China near the Shenzhen city. The Daya Bay nuclear power complex consists of six reactor cores providing a total of up to 17.4 GW thermal power. There are three underground experimental halls (EHs), two near halls and one far hall. For near halls, each hall contains two detectors for determining the reactor neutrino flux; for far site, there are four detectors for measuring the neutrino oscillations. The baselines had been optimized.

\begin{figure}[htb]
\centering
\subfigure{
\includegraphics[height=2.5in]{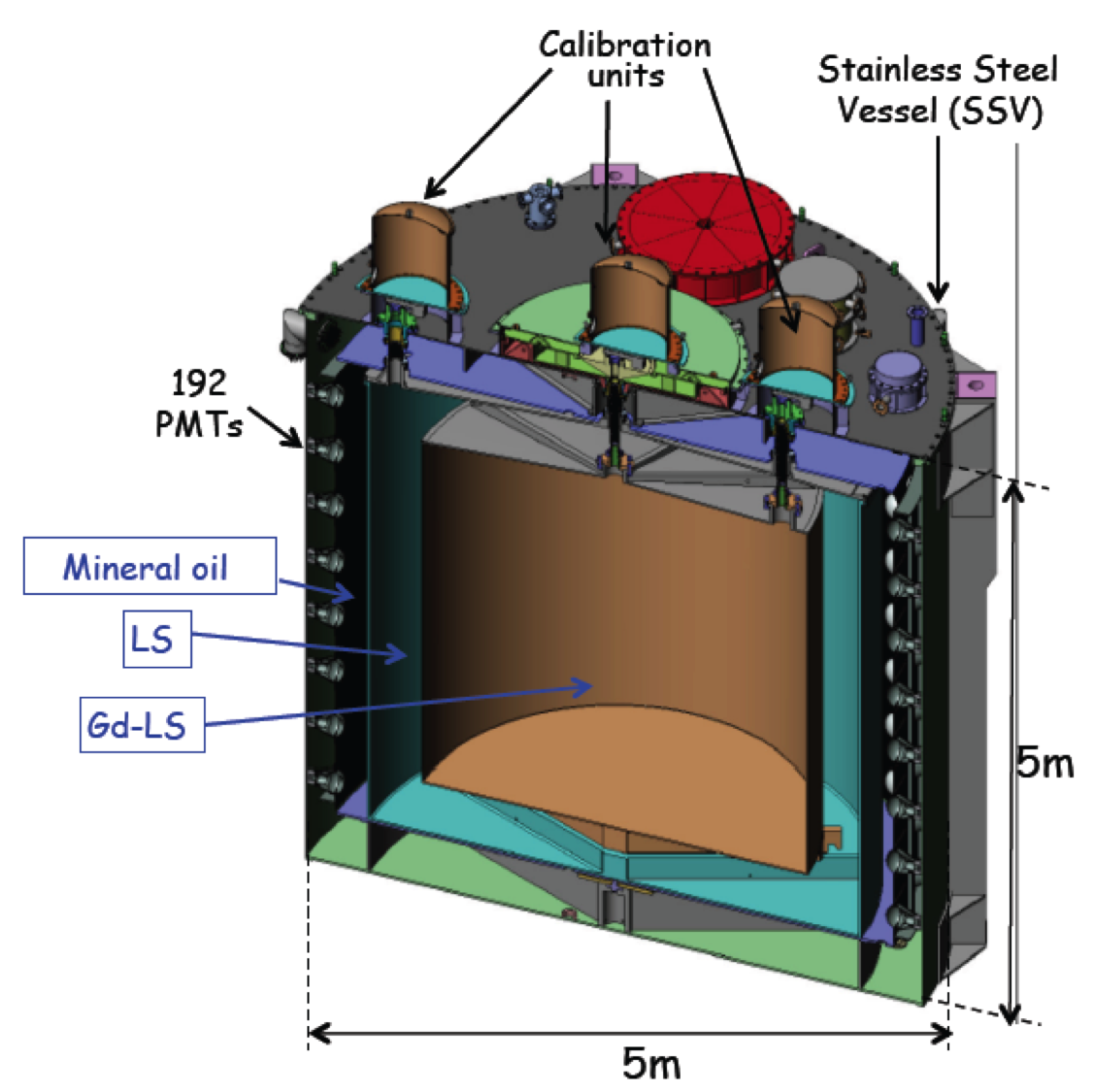}}
\hspace{0.5 in}
\subfigure{
\includegraphics[height=2.5in]{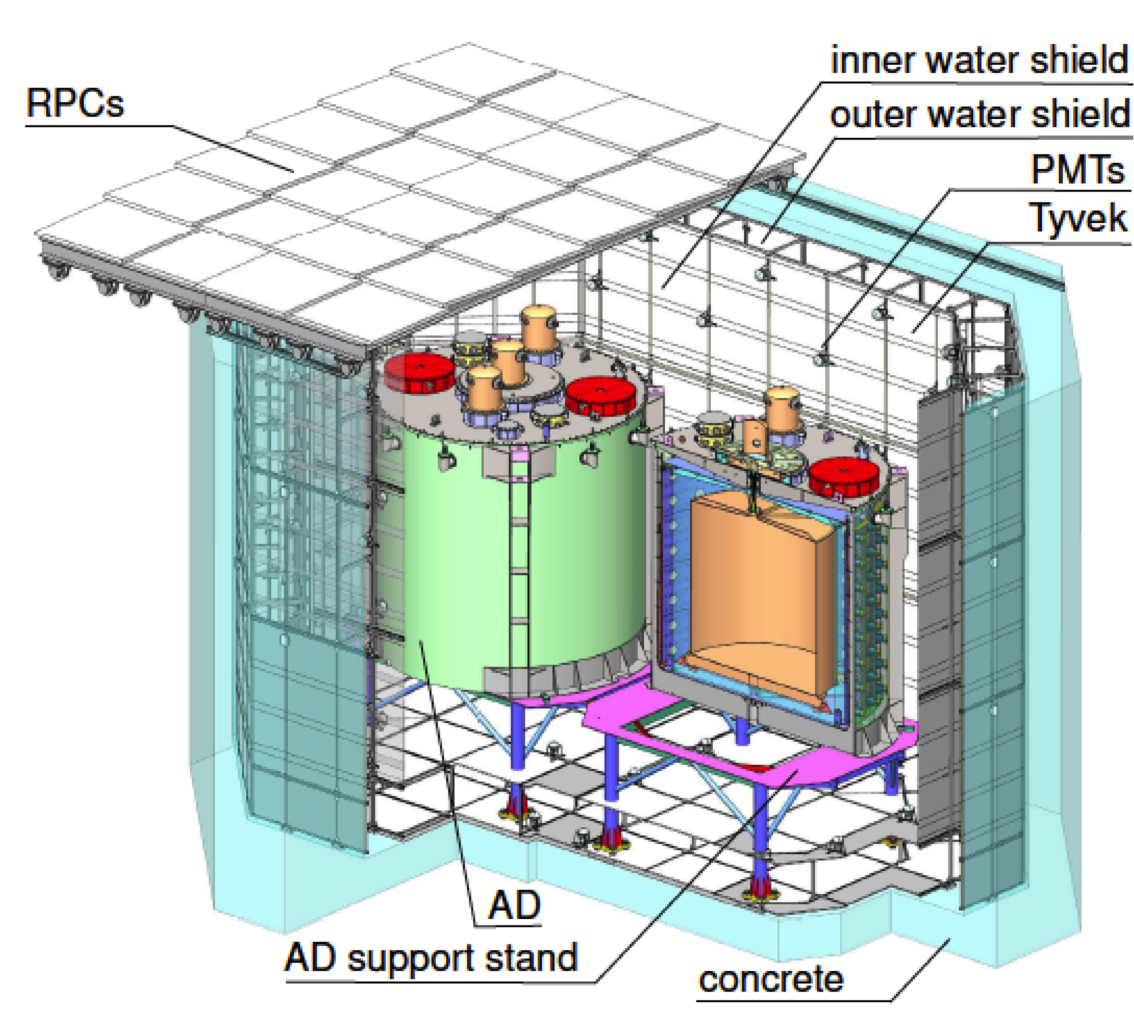}}
\caption{Left: The cross-sectional view of an antineutrino detector. Right: Schematic of a near site experimental hall.}
\label{fig:detectors}
\end{figure}

The antineutrino detectors (ADs) are functionally-identical. Each one consists of three zones. Figure~\ref{fig:detectors} (left panel) is a cross-sectional view of an AD. The inner zone is the antineutrino target which contains 20 tons of Gd-doped liquid scintillator. The middle zone contains 20 tons of liquid scintillator for detecting the gammas escaping from the target volume. The outer zone contains 40 tons of minimal oil for shielding the radioactive background. There are 192 PMTs in each AD and three automatic calibration units (ACUs) on the top of the detector. Each ACU has an LED and three sources ( $^{68}Ge$, $^{60}Co$ and $^{241}Am-^{13}C$). 

The muon tagging system consists of the water pool and the four-layer RPCs. The antineutrino detectors are merged in the water pools. The efficiency of the muon tagging is greater than 99$\%$. Figure~\ref{fig:detectors} (right panel) is a schematic of a near site experimental hall. 

\section{Analysis}
\subsection{Event Selection}
First of all, the instrumental background from the spontaneous light emission by PMTs ( ``flasher" ) is removed from the data. We then remove the muon background. The IBD candidates are selected by the prompt-delayed coincident signals. The prompt signals are between 0.7 MeV and 12 MeV and the delayed signals are between 6 MeV and 12 MeV, while the time separation of two signals is between 1 $\mu s$ and 200 $\mu s$. Finally, we apply the multiplicity cut to remove the ambiguities in the IBD pair selection. One of the multiplicity cut used in our analysis requires no additional prompt-like signals 400 $\mu s$ before the delayed event, and no delayed-like signals 200 $\mu s$ after the delayed event. The other produces consistent result.

\subsection{Backgrounds}

We consider five sources of background. In the antineutrino sample, the accidental background is the largest one. The rate and the spectrum of the accidental background have been determined by measuring the singles rates of prompt- and delayed-like signals. It gives 0.3$\%$ relative uncertainty. The other four are all significantly smaller, but they are correlated backgrounds. The fast neutron and $\beta$-n decay of the cosmogenic $^9Li/^8He$ provides the prompt- and delayed-like signals. Besides, the gamma emission from the neutron capture also affects the IBD detections, which can come from the calibration source $^{241}Am-^{13}C$ and the $^{13}C(\alpha, n)^{16}O$ background.

\subsection{Energy Response}
The energy response is non-linear between the particle true energy and the reconstructed energy. This non-linearity relation, $f$, is caused by both scintillator and electronics effects; the former is related to the quenching effect and Cherenkov light emission, the latter is related to the charge collection of the front-end electronics. The energy response model is defined as $f = f_{scint} \times f_{elec}$. 

For electrons, the empirical model of scintillator nonlinearity is described by $ f_{scint}(E_{true}) = E_{vis}/E_{true} = (p_0 + p_3 \cdot E_{true})/(1+ p_1 \cdot e^{ - p_2 \cdot E_{true} } ) $. The response models of gamma and positrons are connected to the electron model through the Monte-Carlo method. The electron non-linearity model is an empirical exponential function. There exist different energy response models based on different methodologies. The details are given in \cite{dybArxiv}.

The gamma sources and $^{12}B$ spectrum are used to constrain the non-linearity parameters. Finally, positron energy response models obtained from different methods are consistent with each other to $\sim1.5\%$. 

\begin{figure}[htb]
\centering
\subfigure{
\includegraphics[height=1.8in]{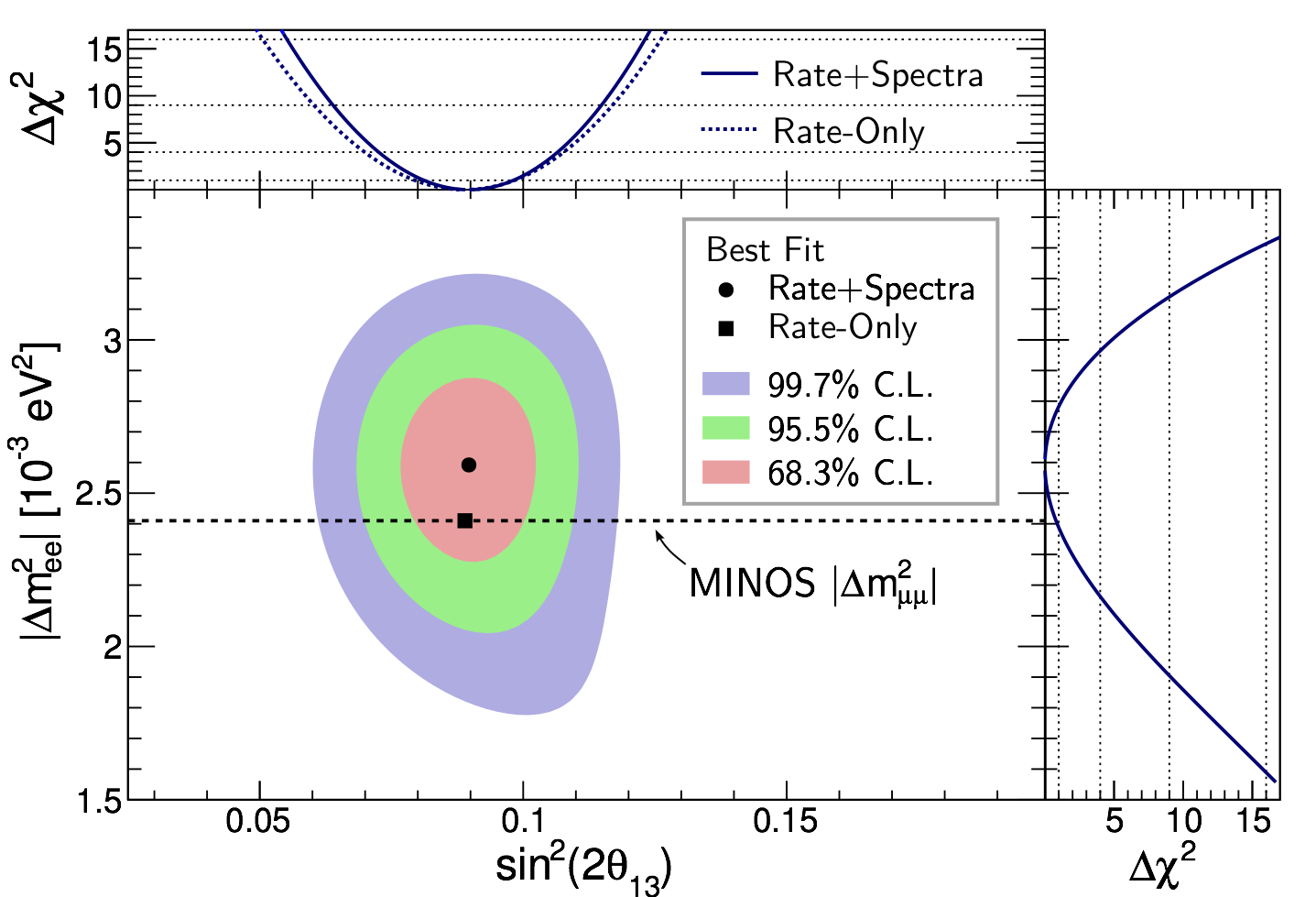}}
\hspace{0.1 in}
\subfigure{
\includegraphics[height=1.6in]{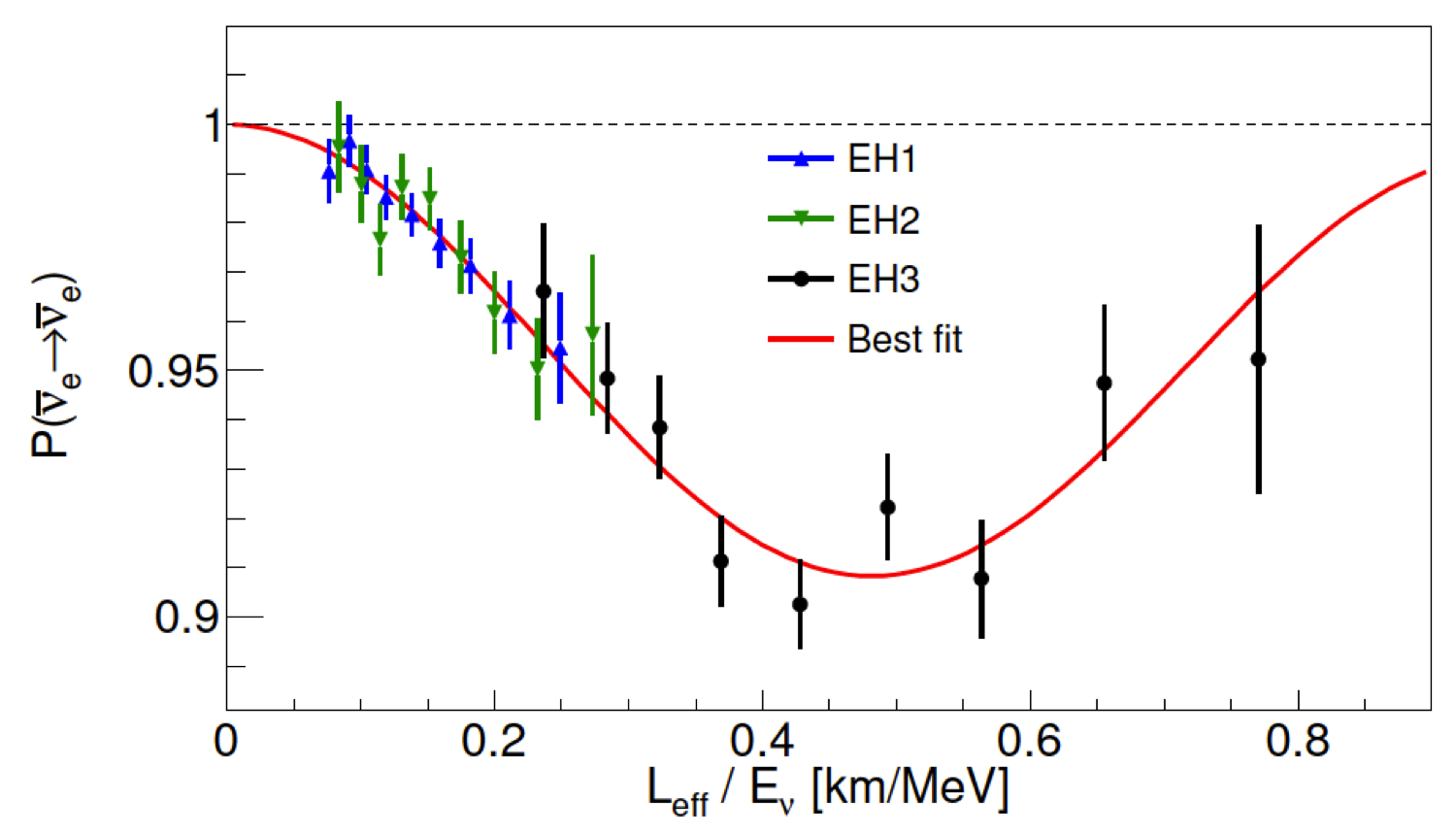}}
\caption{Left: Allowed regions for the $\sin^2 2\theta_{13}$ and $|\Delta m^2_{ee}|$ at the 68.3$\%$, 95.5$\%$ and 99.7$\%$ confidence levels. Right: The survival probability of the electron antineutrino as a function $L_{eff}/E_\nu$, with the best estimate of the detector response. Here $L_{eff}$ is the effective detector-reactor distance, and $E_{\nu}$ is the neutrino energy inferred from the background-subtracted positron energy spectrum.} 
\label{fig:results}
\end{figure}

\section{Conclusion}
The best-fit results for the electron antineutrino oscillation frequency and the value of $\sin^2 2\theta_{13}$ are $| \Delta m^2_{ee}| = 2.59^{+0.19}_{-0.20} \cdot 10^{-3} $eV$^2$ and $\sin^2 2\theta_{13}=0.090^{+0.008}_{-0.009}$ , respectively with a $\chi^2$/NDF of 162.7/153. Under the assumption of normal (inverted) neutrino mass hierarchy, we obtain $| \Delta m^2_{32}| = 2.54^{+0.19}_{-0.20} \cdot 10^{-3} \rm eV^2$ ($| \Delta m^2_{32}| = 2.64^{+0.19}_{-0.20} \cdot 10^{-3} \rm eV^2$). This result is consistent with $| \Delta m^2_{\mu \mu}| $ measured by MINOS \cite{minosPRL}. In Figure~\ref{fig:results}, the left panel shows the allowed regions of 68.3$\%$, 95.5$\%$ and 99.7$\%$ C.L. in the $\sin^2 2\theta_{13}$ vs. $|\Delta m^2_{ee}|$ plane. The right panel shows the comparison of the IBD data with the survival probability by applying the best-fitted values of $\sin^2 2\theta_{13}$ and $|\Delta m^2_{ee}|$ \cite{dybArxiv}.

The total uncertainty is dominated by the statistics. For $\sin^2 2\theta_{13}$, the most significant contributions to the systematic uncertainty are from the reactor, relative detector efficiency and the energy scale. The systematic uncertainty of $|\Delta m^2_{ee}|$ is dominated by the relative energy scale and efficiency. The above results will be improved with the higher statistics and the eight-detector measurement. The measurement of the absolute reactor neutrino flux and other studies taking advantage of Daya Bay detector capabilities will be performed as well.



\Acknowledgements
The Daya Bay experiment is supported in part by the Ministry of Science and Technology of China, the United States
Department of Energy, the Chinese Academy of Sciences, the National Natural Science Foundation of China, the
Guangdong provincial government, the Shenzhen municipal government, the China Guangdong Nuclear Power Group,
Shanghai Laboratory for Particle Physics and Cosmology, the Research Grants Council of the Hong Kong Special
Administrative Region of China, University Development Fund of the University of Hong Kong, the MOE program
for Research of Excellence at National Taiwan University, National Chiao-Tung University, NSC fund support from
Taiwan, the U.S. National Science Foundation, the Alfred P. Sloan Foundation, the Ministry of Education, Youth and
Sports of the Czech Republic, the Czech Science Foundation, and the Joint Institute of Nuclear Research in Dubna,
Russia. We thank Yellow River Engineering Consulting Co., Ltd. and China railway 15th Bureau Group Co., Ltd.
for building the underground laboratory. We are grateful for the ongoing cooperation from the China Guangdong
Nuclear Power Group and China Light $\&$ Power Company.

\end{document}